\documentclass[sigconf]{acmart}

\usepackage{booktabs} 

\usepackage{graphics}
\usepackage{subcaption}
\usepackage{booktabs} 
\usepackage{tikz}
\usepackage{amssymb}
\usepackage{graphicx}
\usepackage{hyperref}
\usepackage{amsmath}
\usepackage{color}
\usepackage{array,booktabs}
\usepackage{algorithm}
\usepackage{algpseudocode}
\usepackage{tabu}

\makeatletter
\def\BState{\State\hskip-\ALG@thistlm}
\newcommand{\mypm}{\mathbin{\mathpalette\@mypm\relax}}
\newcommand{\@mypm}[2]{\ooalign{%
  \raisebox{.1\height}{$#1+$}\cr
  \smash{\raisebox{-.6\height}{$#1-$}}\cr}}
\makeatother

\renewcommand\footnotetextcopyrightpermission[1]{} 

\acmConference[KDD '18]{ACM KDD conference}{August 2018}{London, United Kingdom}

\begin{document}
\title{Murmur Detection Using Parallel Recurrent \& Convolutional Neural Networks}

\author{Shahnawaz Alam, Rohan Banerjee, Soma Bandyopadhyay}
\email{{shahnawaz.alam, rohan.banerjee, soma.bandyopadhyay} @tcs.com}
\affiliation{%
  \institution{TCS Research}
  \city{Kolkata}
  \country{India}}

\renewcommand{\shortauthors}{S. Alam et al.}

\begin{abstract}
In this article, we propose a novel technique for classification of the Murmurs in heart sound. We introduce a novel deep neural network architecture using parallel combination of the Recurrent Neural Network (\begin{math}RNN\end{math}) based Bidirectional Long Short-Term Memory (\begin{math}BiLSTM\end{math}) \& Convolutional Neural Network (\begin{math}CNN\end{math}) to learn visual and time-dependent characteristics of Murmur in PCG waveform. Set of acoustic features are presented to our proposed deep neural network to discriminate between Normal and Murmur class. The proposed method was evaluated on a large dataset using 5-fold cross-validation, resulting in a sensitivity and specificity of 96 $\mypm$ 0.6 \% , 100 $\mypm$ 0 \% respectively and F1 Score of 98 $\mypm$ 0.3 \%.

\end{abstract}

%
%



\maketitle

\section{Introduction}
Cardiovascular diseases (CVDs) continue to be the leading cause of morbidity and mortality worldwide. An estimated 17.5 million people died from CVDs in 2012, representing 31\% of all global deaths (WHO 2015). Cardiac auscultation using stethoscope has been a classic procedure for screening heart abnormalities. Heart sound (Phonocardiogram i.e. PCG) auscultation though being noninvasive and inexpensive to capture, has limitations: dependency on hearing sense to judge unbiasedly, and precise skill which takes years of experience to gain, therefore, are not fully reliable. However, digital signal processing on PCG has shown a high correlation with the evaluation of primary healthcare and cardiac defects, which makes it a popular choice. Therefore, intelligent agent over auscultation would assist medical practitioners with decision support abilities.

Heart sounds consist of two fundamental / primary beats, called S1 \& S2 (sounds like `lub' \& `dub'), occurring sequentially in time domain. A single mechanical cycle of heart, or cardiac cycle, produces an S1 sound when the closing of the mitral \& tricuspid valves occur followed by Systole then an S2 sound by the closing of the pulmonary \& aortic valves followed by Diastole. This ideal blood flow in the heart is streamlined; thus interbeat sounds are minimal; however, turbulence in the blood flow due to structural defects and other diseases in the heart creates vibration in the surrounding tissue, thus creating audible noise, called pathologic murmur. Murmurs have diverse and wider frequency ranges of sound compared to primary heart sound (S1 \& S2) and depending on their nature they can be as high as 600 Hz. Murmurs are one of the pathological indicators of abnormalities in the heart, which needs medical attention. Functional, or innocent murmurs are caused due to physiologic conditions outside the heart, which are benign and not included in this study.  

Murmurs, depending upon the position in the cardiac cycle can be further classified into systolic and diastolic murmurs; however, they quickly get mixed up with cardiac beats which makes it difficult to identify by our hearing sense. Therefore, we need an automatic classifier which perceives inherent features of PCG to discriminate normal and murmur heart sound. 

In recent times, PCG murmur detection has been an active area of research, with various studies centering on the determination of a multitude of features extraction which correlates to murmur, mostly related to time, frequency, morphological, parametric modeling and spectral properties of heart sounds. 
In \cite{liang1998feature}, Liang et al. have extracted features based on coefficients of wavelet packet decomposition to classify heart sound into physiological and pathological murmurs. Features were introduced to a neural network to get 85\% accuracy. However, it does not provide information about the normal class or noisy recordings which might have a similar pattern in wavelet space, thus difficult to segregate.
In \cite{sharif2000analysis}, Sharif et al. have taken instantaneous energy and frequency estimation (central finite difference frequency estimation and zero crossing frequency estimation) of heart sound for the classification. 
In \cite{ari2010detection}, Ari et al. have extracted wavelet-based features to discriminate normal and abnormal heart sound using modified Support Vector Machine (Least Square SVM) - which introduces Lagrange multiplier based on Least Mean Square algorithm. 
In \cite{delgado2009digital}, Delgado et al. have used perceptual \& fractal features to understand the temporal patterns of the heart sound murmur, in which fractals have shown significant performance boost over other. 
Yoshida et al. \cite{yoshida1997instantaneous} have analyzed the systolic duration of heart sound using instantaneous frequency from averaged Wigner-Ville distribution for the characterization of PCG. 
In \cite{rubin2017recognizing}, Rubin et al. have used deep learning approach to classify the time-frequency heat maps using a convolutional neural network (CNN) which resulted into 0.73 sensitivity and 0.95 specificity on Physionet Challenge 2016 \footnote{\url{https://www.physionet.org/challenge/2016/}} dataset.

Previous studies have mostly focused on the feature extraction based method for classification. However, handcrafted features mostly fail in the on-field test environment, as these features are highly biased towards training dataset. Moreover, number of the prior studies for heart sound classification is flawed because of utilization of carefully-selected data, failure to use a variety of PCG recordings from different demographics and pathological conditions, training and evaluation on clean recordings, etc. Few studies have also used heart sound segmentation to extract features, which is a difficult task on recordings having murmur even with the current state-of-the-art technique \cite{springer2016logistic}. Segmentation can fail miserably due to the presence of noise, artifacts and other sounds including third heart sound `S3', fourth heart sound `S4', and murmur.

Heart sounds are inherently prone to interfering noise (ambient, speech, etc.) and motion artifact, which can overlap time location \& frequency spectra of murmur in heart sound. Mumur being mostly pseudo-periodic in nature (assuming noises are random and varying), spectral and temporal sequence analysis best suits the attempt for classification which adapts and tracks the statistical property of the signal over a time-frequency space. Recent advancements in deep learning have shown a promising result; therefore we utilize Convolutional \& Recurrent Neural Network to solve the classification problem.

The rest of the paper is organized as follows: Section 2 describes the datasets, pre-processing and feature engineering, the architecture of the proposed network is detailed in Section 3. Section 4 describes the training procedure. Results and observation are explained in section 5, and finally, section 6 concludes the article.

\section{Dataset Description}
We have used three publicly available heart sound datasets including 
\begin{enumerate}
\item D1 : `Heart Sound \& Murmur Library' \footnote{\url{http://www.med.umich.edu/lrc/psb_open/html/repo/primer_heartsound/primer_heartsound.html}} by University of Michigan
\item D2 : `Classifying Heart Sounds Challenge' \footnote{\url{http://www.peterjbentley.com/heartchallenge/}} by Pattern Analysis, Statistical Modelling and Computational Learning (PASCAL).
\item D3 : `Physionet Challenge' 2016 dataset\footnote{\url{https://www.physionet.org/challenge/2016/}} by Computing in Cardiology
\end{enumerate}

Dataset D1 contains in total 23 recordings having various abnormal heart sound out of which 14 were murmur recordings while remaining 9 were the normal category. D2 have in total 585 annotated recordings, out of which 129 are labeled as murmurs (which includes 29 noisy data). Remaining 456 recordings (with 120 noisy data) are labeled normal heart sound which includes recording with artifacts, extrasystole, extra heart sound (S3, S4, and Clicks) and normal. To further enhance the dataset, we have considered D3 with 3240 data (normal and abnormal) collected from patients around the world having various pathological conditions. However, we have considered only normal class 2575 data (including 186 noisy) and have discarded abnormal (665 recordings) heart sound from D3. As an abnormal class in this dataset includes patients recordings having both heart valve defects (including murmur sound) and coronary artery disease (CAD), and Physionet does not provide sub-classification for these recordings. 

\begin{figure}[h]
    \centering
    \includegraphics[width=0.5\textwidth]{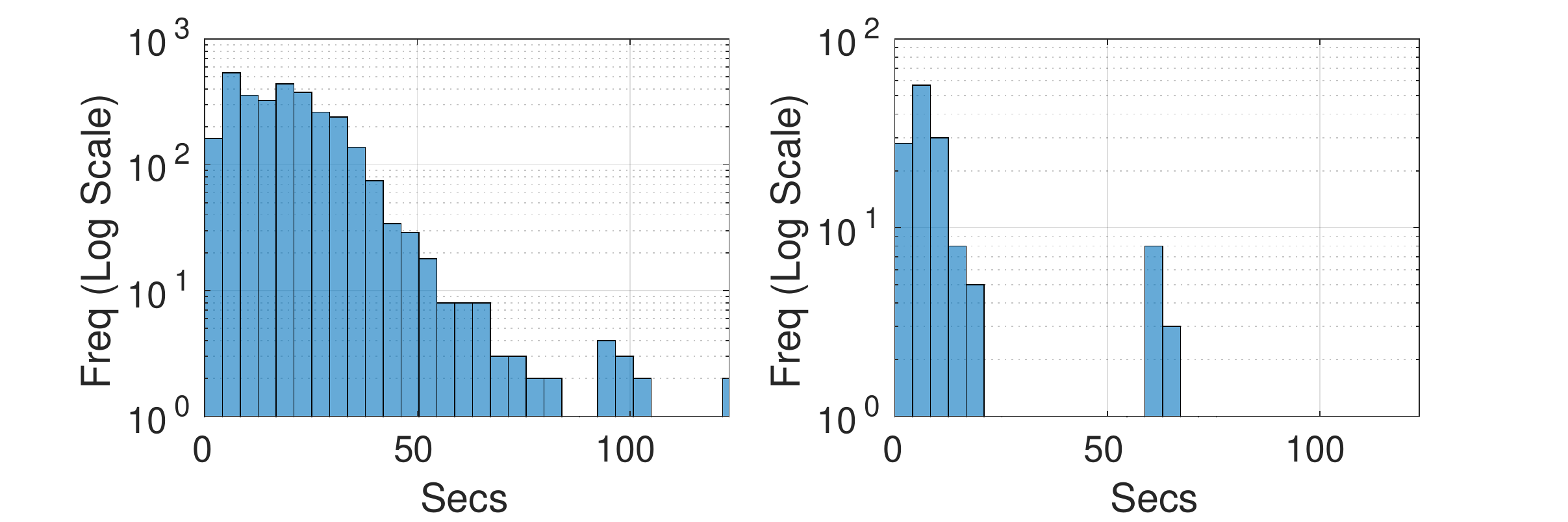}
    \caption{Histogram of length of recordings of Normal and Murmur class}
    \label{fig:length}
\end{figure}

Therefore, combining above datasets, we create a single pool of data having 143 murmur recordings (14 from D1 \& 129 from D2) \& 3040 normal recordings (9 from D1, 456 from D2 \& 2575 from D3). The sampling frequency of dataset was maintained at 2000 Hz. The data length of recordings ranges from less than a second to just above 123 seconds as shown in the figure \ref{fig:length}. 

We have created a challenging dataset in terms of sensor diversity, different demographics, various location of recordings, different pathological conditions (Physionet data were collected from ICU/CCU), the variation of age \& BMI of subjects, variety of noises (ambiance, motion artifacts), varying sampling frequencies, and class imbalance. Therefore, we pursue a generalized classifier learner over the dataset, where deep filters and long-term dependency aggregated by our Deep Neural Network (DNN) architecture learns the highly correlated features with class and are sensor and noise independent. 

\subsection{Data Segmentation}
Due to variation in the length of recordings, sequences are segmented into non-overlapping chunks of 4 sec length each. The analogy behind this duration is to focus on the minute and deep feature to understand murmur rather than learning very long-term dependency which is independent of our study (e.g., Heart Rate Variability and breathing effect). Residual sequences of very short length (< 2 sec) are not guaranteed to provide sufficient murmur information and are thus removed from the data pool, whereas others are padded with trailing zero. After segmentation, we have 10,892 instances of data out of which 272 belongs to murmur class (class ratio nearly equals 39:1 normal:murmur).

\subsection{Feature Engineering}
Deep learning has provided many tools for automatic feature extraction based on raw data and its annotation, for example, Autoencoder, and Convolutional Neural Network kernels on raw signals to learn important filters. However, these work best for the spatial feature extraction of significantly high-resolution signal compared to our single dimension heart sound. Therefore, in this study, we extract features for the classification of signals.

Murmur, being a pseudo-periodic non-stationary high-frequency signal, we need both temporal and spatial trend to discriminate the disease from normal class. This will also reduce biases towards noises (ambiance and motion artifact) which are relatively aperiodic and of varying frequency spectra. 

Before feature extraction, we apply low pass followed by high pass filters with cutoff frequencies 500Hz and 25Hz respectively. We have also removed noisy spike in the signal using \cite{schmidt2010segmentation} method. We utilize acoustic features: Mel filter based Mel-Frequency Cepstral Coefficients (MFCC), a well-known technique in speech processing, and short-term Fourier transform based spectrogram to find the time-dependent and visual signature of the signal respectively. We use the following method to compute these features :

\subsubsection{Spectrogram}

\begin{enumerate}
\item Signals are divided into small windows of length 128ms and step size of 64ms.
\item On each window, Fast Fourier Transform (FFT) is applied with a Hamming window of the length of 128.
\item This resulted to spectrogram image of dimension [65 x 61] for a 4 sec heart sound. 
\end{enumerate}

\subsubsection{Cepstrogram (MFCC)}

\begin{enumerate}
\item Signals are divided into small windows (with window index \begin{math}i\end{math}) of length 25ms and step size of 10ms.
\item On each window Discrete Fourier Transform (DFT) \begin{math}D_{i}(k)\end{math} (\begin{math}k \in [1,K]\end{math}, K is length of DFT) are applied with Hamming window of length of 50.
\item Spectral power estimate \begin{math}P_{i}(k)\end{math} is computed on the DFT signal for every window.
\item 26 Filterbanks channels of triangular band-pass filters (with cutoff frequency range \begin{math}[70,500]\end{math}) followed by log transformation is applied to \begin{math}P_{i}(k)\end{math}.
\item Discrete Cosine Transform is finally applied to compute MFCCs with 13 coefficients for each window \begin{math}i\end{math}. 
\item Cepstral feature of each window are combined to get Cepstrogram sequence of dimension [13 x 398] for a 4 sec heart sound. 
\end{enumerate}

\section{Architecture}
\begin{figure}[h]
    \centering
    \includegraphics[width=0.4\textwidth]{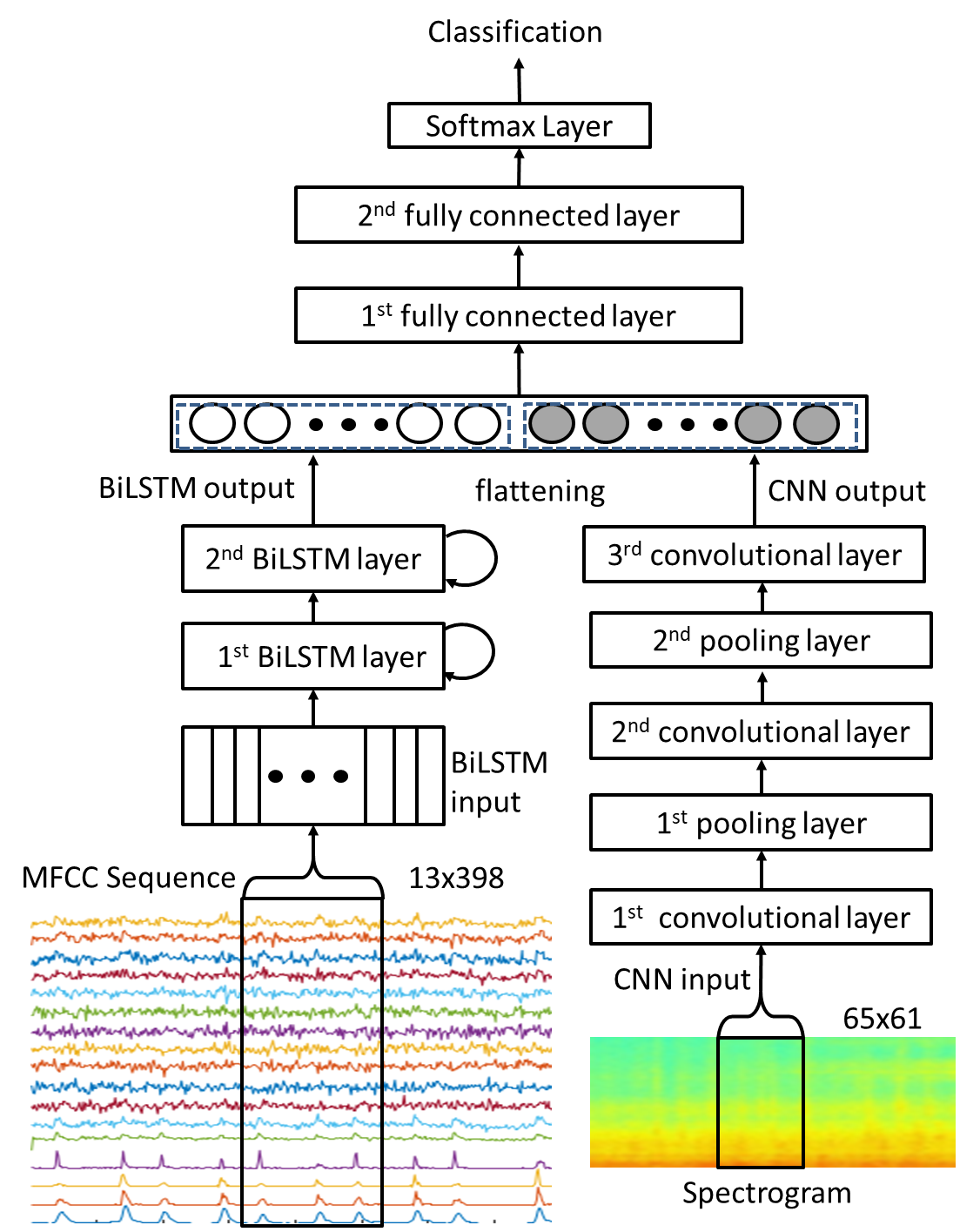}
    \caption{Deep Neural Network Architecture}
    \label{fig:Architecture}
\end{figure}

In this study, we propose two parallel deep neural networks: Recurrent Neural Network based Bidirectional Long Short-Term Memory (\begin{math}BiLSTM\end{math}) and Convolutional Neural Network (\begin{math}CNN\end{math}) which merges by flattening and concatenating their respective output following fully connected layers with common loss back-propagation path. The LSTM layers utilize the sequence of acoustic MFCCs feature, while the CNN layers use spectral images as input. The schema of the proposed neural network is shown in figure \ref{fig:Architecture}. 

\subsection{CNN}
Convolutional layer performs 2-dimensional convolution between the spectral image and the trained filters. To learn the different aspect of the features, numbers of filter channels are used. Thus, when we apply N different filters to the spectrogram, N output filtered images \begin{math}F_{i}\end{math} (\begin{math}i \in [1,N]\end{math}) are computed in the convolutional layer. The filtered images \begin{math}F_{i}\end{math} are forwarded to the pooling layer which evaluates the sub-segments of \begin{math}F_{i}\end{math} and compute maximum value which downsamples the image. This spatial downsampling ensures the most dominant feature in the sub-region is extracted. In the final pooling layer, the resized outputs are flattened in order to connect with the subsequent fully connected layers.
 
\subsection{BiLSTM}
LSTM (or RNN) contains hidden layers with self-recurrent weights which enables the cell (nodes) in the memory block to retain past information. This inference of prior information is used for future estimation; thus LSTM is famous for extracting the temporal pattern of the sequence. Bidirectional LSTM (\begin{math}BiLSTM\end{math}) is the modified LSTM which has a bidirectional flow to process the sequence in both forward and backward direction and fed forward to the output layer. Two hidden layers present in \begin{math}BiLSTM\end{math} computes hidden sequences both in forward and backward direction and updates the output layer by using backward layer (from last time step to the first) and forward layer (from first to last time step). In our system, \begin{math}BiLSTM\end{math} learns the temporal trend of MFCC's sequences of the heart sound. 

\section{Training}
To train our network, we need to feed two sets of features (Spectrogram \& MFCC sequence) of every instance simultaneously in the same order in every iteration of the epoch. Therefore, we maintain two ordered datasets for each corresponding features accordingly. For robust performance evaluation, we do 5-fold cross-validation over 10,892 data. However, due to the predominance of class imbalance (nearly 39:1 normal:murmur as quoted earlier), training would be bias towards majority class. Therefore, we augment the minor class in training data in each fold independently by upsampling (repeating minor class) to create balance class training set, however, we have left test data in every fold as it is. Moreover, we also made sure that data segments of the same patient (or subject) are not present in both train and test in any of the folds. This insures non over-optimistic result. 

\subsection{CNN}
Proposed CNN architecture comprises three convolutional layers, two max-pooling layers, and a fully connected layer to flatten the output of \begin{math}CNN\end{math}. The input to this \begin{math}CNN\end{math} is 65x61 spectrogram. In the first convolutional layer, the input is convolved with 4 filters of size 3x3. Batch normalization followed by ReLU activation function was applied on the output of the convolutional filter. First max-pooling layer summarizes and reduces the size of filters using 2x2 kernel. Similarly, two subsequent convolutional layers convolve output of max-pooling layers using 3x3 filter followed by batch normalization and ReLU activations. Final activation output is then flattened and feed to fully connected layer with 128 units. To reduce over-fitting, we used L2-Regularization over all the layers in CNN.
 
\subsection{BiLSTM}
In the \begin{math}BiLSTM\end{math} architecture, input sample consist 398 frames with 13-dimensional MFCC features. The network have two \begin{math}BiLSTM\end{math} layers with 128 \begin{math}BiLSTM\end{math} units each and one feed-forward layer with 128 ReLU units similar to CNN output. Dropout was applied to all the layers with keeping probability of 80\%.

We built parallel \begin{math}BiLSTM\end{math} and \begin{math}CNN\end{math} structure and combine their outputs (128 each) to form a single fully connected layer with 256 units. Second fully connected layer has 128 units which is connected to output layer containing two softmax nodes identical to the number of class (Normal \& Murmur). The network structure and flow of \begin{math}BiLSTM\end{math} and \begin{math}CNN\end{math} part is portrait in above sub-sections. This combined structure was then trained over the dataset explained above. 

All the networks were trained using cross-entropy error as the loss function. In each epoch, we set 128 mini-batches in random order (however with same seed over two feature sets to maintain uniformity in order). Learning rate was kept 1E-3 throughout training.  

\section{Results and Observations}
We present a set of performances measures (Sensitivity (the portion of normal class predicted normal), Specificity (the portion of murmur class predicted murmur), \& F1 Score of Normal class) over 5-fold cross-validation as shown in Table \ref{result}. We have provided performance of our network (\begin{math}BiLSTM + CNN\end{math}) and individual \begin{math}CNN\end{math} \& \begin{math}BiLSTM\end{math} networks too which were trained \& tested independently. As we can see \begin{math}CNN\end{math} \& \begin{math}BiLSTM\end{math} networks are bit biased towards Murmur and Normal class respectively. This is because \begin{math}CNN\end{math} network learns visual filters for discrimination; however, it faces difficulties differentiating noisy data of Normal class and Murmur. On the other hand, \begin{math}BiLSTM\end{math} learns the long-term repetitive patterns of the principle heart sound (S1, \& S2) of Normal class but fails to discriminate few instances of Murmur sound which are dominant in periodicity. Therefore, when both networks are combined (\begin{math}BiLSTM + CNN\end{math}) in parallel, we observe the network learns visual as well as the time-dependent aspect of the signal and are a better place to discriminate the classes. 

\renewcommand{\tabcolsep}{4pt}
\begin{table}[h]
\centering
\caption{5-fold cross-validation results}
\label{result}
\begin{tabular}{|l|l|l|l|}
\hline
                      & \textbf{Sensitivity}      & \textbf{Specificity} & \textbf{F1 Score}         \\ \hline
\begin{math}CNN\end{math}                   & 0.8704 $\mypm$ 0.0137          & 1 $\mypm$ 0               & 0.9307 $\mypm$ 0.0079          \\ \hline
\begin{math}BiLSTM\end{math}                & 0.9714 $\mypm$ 0.0079          & 0.8549 $\mypm$ 0.0636     & 0.9835 $\mypm$ 0.0042          \\ \hline
\begin{tabular}{@{}c@{}}\begin{math}CNN\end{math} \\ \begin{math}+\end{math} \\ \begin{math}BiLSTM\end{math}\end{tabular} & \textbf{0.9609 $\mypm$ 0.0061} & \textbf{1 $\mypm$ 0}      & \textbf{0.9801 $\mypm$ 0.0032} \\ \hline
\end{tabular}
\end{table}

The result shows we are achieving 100\% Specificity \& 96\% Sensitivity, a significant boost from prior arts \cite{sharif2000analysis} \cite{delgado2009digital}. The major positive point of our proposed system lies behind the fact that it does not depends on beat (heart cycle from an S1 to next occurring S1) segmentation algorithm, which has its own sensitivity and specificity, and is independent of onset estimation.  

\section{Conclusions}
In this study, we propose a deep neural network solution to classify Normal and Murmur heart sound signals automatically. The proposed method is trained \& tested using internal 5-fold cross-validation on a large corpus of a diverse dataset comprising of multiple sensor PCG signals with varying degrees of noise, motion artifacts \& pathological conditions. Our system learns the visual and time-dependent properties of the heart sound signal using parallel networks of \begin{math}BiLSTM\end{math} \& \begin{math}CNN\end{math}. The results demonstrate that automated Murmur detection of PCG signal using acoustic MFCC features and Spectrogram trained on our proposed network leads to high performance.

\bibliographystyle{ACM-Reference-Format}
\bibliography{sample-bibliography}

\end{document}